\documentclass[twocolumn,prl]{revtex4}
\usepackage[dvips]{graphicx}
\usepackage{amsfonts}
\usepackage{amsmath}
\usepackage{comment}
\usepackage{float, color,array, multirow}

\begin{document}
\title{Fractal and subharmonic responses driven by surface acoustic waves\\during charge density wave sliding}
\author{Yu Funami and Kazushi Aoyama}
\date{\today}
\affiliation{Department of Earth and Space Science, Graduate School of Science, Osaka University, Osaka 560-0043, Japan}

\begin{abstract}
We theoretically investigate the effects of surface acoustic waves (SAWs) on an electric-field-driven sliding motion of a one-dimensional charge density wave (CDW), which is initially pinned by impurities. By numerically analyzing an extended Fukuyama-Lee-Rice model, we show that a mechanical vibration of the SAW, which, in the model, is assumed to affect the CDW via the pinning site in the form of temporally oscillating pinning parameters, induces Shapiro steps with self-similarity, i.e., the devil's staircase, in the current-voltage characteristics. It is also found that when the SAW acts as the vibration in the pinning strength, the mechanism of the mode locking (harmonic and subharmonic responses) leading to the occurrence of the Shapiro steps is modified, and as a result, the fractal dimension and parameter dependence of the SAW-induced staircase can be considerably  different from those for the conventional ac-electric-field-induced one. This suggests that an unconventional type of fractal phenomena can emerge in the SAW-induced CDW dynamics.
\end{abstract}

\maketitle
Fractals with their characteristic properties exemplified by self-similarity and non-integer dimensions often appear in nature in various forms such as coastlines, snowflakes, and polymer chains \cite{Fractal_review, Fractal_polymer_experiment}. Of particular interest are fractal phenomena induced by dynamical effects, one example of which is a sliding motion of a charge density wave (CDW), an electron condensate emerging typically in quasi-one-dimensional conductors such as $\mathrm{NbSe}_3$, $\mathrm{TaS}_3$, and $\mathrm{K_{0.3}{MoO}_{3}}$ \cite{C0018,C0070,C0098}.
When the one-dimensional CDW is driven to slide by an external dc electric field $E_{\rm dc}$ \cite{Frohlich_CDW,C0029,C0086}, an electric current carried by the sliding CDW $I_{\rm CDW}$ as a function of $E_{\rm dc}$, i.e., the $I$-$V$ characteristics, is known to exhibit step-like fractal structures, the so-called devil's staircase, in the presence of an additional ac electric field $E_{\rm ac}$ \cite{C0014,C0036_C0017,C0017_2,DS}.
Subharmonic responses emerging in the staircase as fractional-value plateaus are closely related to time crystals in periodically-driven systems \cite{Wilczek_review_arXiv_23}.
In this Letter, motivated by recent experiments where mechanical vibrations \cite{C0056} including a surface acoustic wave (SAW) \cite{private} were applied instead of $E_{\rm ac}$, we theoretically investigate vibration effects on the staircase formation in the CDW sliding.

The CDW state is characterized by a spatial modulation in the electron density which, in one dimension, takes the following form:
\begin{equation}\label{charge-density}
\rho(x,t)=\rho_0+\rho_1 \cos \left(\phi(x,t)+Qx\right).
\end{equation}
Here, $\rho_0$ is the average electron density, and $\rho_1$ and $Q$ are, respectively, the amplitude and wave number of the CDW modulation whose phase $\phi(x,t)$ plays an important role for the CDW dynamics. In materials, the CDW modulation is locally deformed due to impurities, defects, and lattice distortions, which can be described as a pinning of the CDW phase $\phi$ \cite{C0074,C0073}. 
Such a pinned object can be driven by an external force overcoming an associated static friction \cite{C0094,C0096} which in the present CDW case, is an external electric field $E$.
When the dc component of $E$, $E_{\mathrm{dc}}$, exceeds a threshold value, the CDW is depinned and begins to slide with velocity $v=\frac{1}{Q}\frac{d\phi}{dt}$, carrying the associated current $I_{\mathrm{CDW}}$ proportional to $v=\frac{1}{Q}\frac{d\phi}{dt}$ \cite{C0018,C0070,Frohlich_CDW,C0029,C0086}. In this sliding regime, there exists a characteristic frequency $\omega_{\mathrm{\phi}}$, the so-called narrow-band noise \cite{C0005,C0030_C0044}, which corresponds to a period for a specific part of the CDW, e.g., the peak position of the wave, to pass through a fixed pinning site, and thus is given by $\omega_{\mathrm{\phi}} = vQ =\frac{d \phi}{dt}$. This oscillating mode $\omega_{\mathrm{\phi}}$ is related to $I_{\mathrm{CDW}}$ via
\begin{equation}\label{CDW-current}
I_{\mathrm{CDW}}\propto \frac{d\phi}{dt} = \omega_{\mathrm{\phi}}.
\end{equation}
In the case without the ac component ($E_{\rm ac}=0$), $I_{\rm CDW}$, or equivalently, $\omega_{\mathrm{\phi}}$, gradually increases with increasing the driving force $E_{\rm dc}$.

When the dc and ac fields are simultaneously applied, i.e., $E=E_{\mathrm{dc}}+E_{\mathrm{ac}} \sin \left(\omega_{\mathrm{ex}} t\right)$, the sliding mode $\omega_{\mathrm{\phi}}$ is coupled to the external frequency $\omega_{\mathrm{ex}}$, leading to the emergence of plateau regions in the $I$-$V$ characteristics \cite{C0014,C0036_C0017,C0017_2,C0051,C0025,C0003,C0024,C0001,C0022,C0028,Matsukawa_JJAP_1987,C0040_C0041_C0042,C0097,C0071,C0052}. A typical theoretical result is shown in Fig. \ref{fig:I-V} (a). The plateaus where $\omega_{\mathrm{\phi}}$ is mode-locked to $\omega_{\mathrm{\phi}}=(p/q)\omega_{\mathrm{ex}}$ with integers $p$ and $q$ are CDW analog of the Shapiro steps discussed in the context of superconductivity \cite{C0083,C0084,C0085}. The plateaus with integer values of $p/q$ are called harmonic steps and others are called subharmonic steps. In the overdamped regime, the former can be explained by a single-impurity model, whereas the latter corresponding to discrete-time-crystal states \cite{Wilczek_review_arXiv_23} can be explained by many-body multi-impurity models \cite{C0018}. One can see from Fig. \ref{fig:I-V} (a) that so many subharmonic steps construct a self-similar structure in the $I$-$V$ characteristics. Although in general, the occurrences of the subharmonic steps and the devil's staircase are not equivalent, the fractal nature has been confirmed in theoretical works \cite{C0018, C0017_2,C0022,C0065} and in one experiment as well \cite{C0014}. The $E_{\rm ac}$-driven staircase is considered to belong to the universality class of the circle map where the fractal dimension is $D=0.87$ \cite{C0036_C0017,DS}.

\begin{figure}[t]
\begin{center}
\includegraphics[width=\columnwidth]{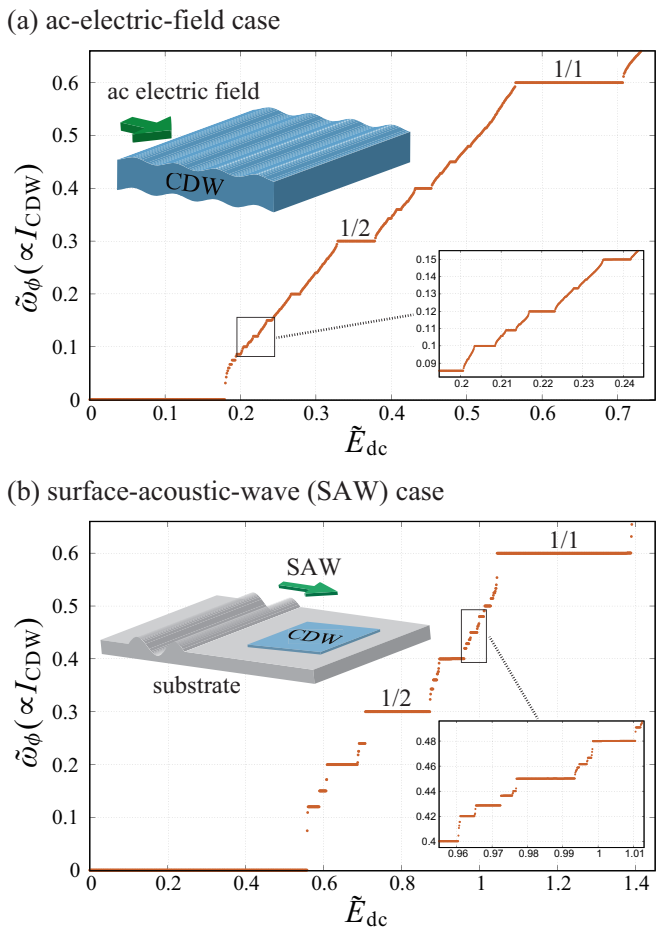}
\caption{The $\,\tilde{E}_{\mathrm{dc}}$ dependence of $\tilde{\omega}_{\phi}\,$ (the $I$-$V$ characteristics) in (a) the ac-electric-field case of $\tilde{E}_{\mathrm{ac}} = 3.0$, $\tilde{P}_{\mathrm{dc}} = 2.0$, and $\tilde{P}_{\mathrm{ac}} = 0$ and (b) the SAW case of $\tilde{E}_{\mathrm{ac}} = 0$, $\tilde{P}_{\mathrm{dc}} = 2.0$, and $\tilde{P}_{\mathrm{ac}} = 1.2$, where the results are obtained for $\tilde{\omega}_{\mathrm{ex}}=0.6$ and $N_{\mathrm{imp}} = 200$. In each figure, the inset shows a magnified view of a small area enclosed by a box, and the top left image shows the system setup in each case.}
\label{fig:I-V}
\end{center}
\end{figure}

Quite recently, the Shapiro steps have also been observed in a different kind of experiment where instead of $E_{\rm ac}$, a time-dependent strain \cite{C0056} was applied. In addition, the effect of another type of mechanical force, the SAW, on CDW sliding has also been reported \cite{private}, though in this case, an electrical contribution as well as the mechanical one might be relevant. In contrast to the electromagnetic interaction between $E_{\rm ac}$ and CDW, how the mechanical forces affect CDW is an interesting issue. In this work, assuming that the mechanical vibration indirectly interacts with the CDW via the pinning sites, we theoretically investigate its effect on the CDW dynamics. We mainly focus on the SAW case with presumably spatially non-uniform strains as a typical platform for a pinning-strength vibration (see Sec. I in \cite{Supplemental_Material}) which is a key ingredient for the following results. We will show that the pinning-strength vibration caused by the SAW induces Shapiro steps with a non-trivial fractal dimension and a parameter dependence of the step width which is qualitatively different from the $E_{\rm ac}$-induced one. 

Although even the threshold-field physics cannot fully be explained by a compact theory \cite{Et_review_Throne_05}, many aspects of the CDW sliding can be described by the Fukuyama-Lee-Rice model \cite{C0057,C0034,C0035} whose overdamped equation of motion in one dimension is given by
\begin{equation}\label{eq:FLR-model}
\begin{aligned}
\left(\gamma\frac{\partial}{\partial t}-v_{\mathrm{ph}}^2\nabla^2\right) \phi=P_{\mathrm{imp}} N_{\mathrm{p}}(x)\sin (\phi+Qx)+\frac{eQ}{m^*} E,
\end{aligned}
\end{equation}
where $e$, $m^{\ast}$, and $v_{\mathrm{ph}}$ are the electric charge, the effective mass of the CDW, and the phason velocity, respectively, and $\gamma$ is the phenomenologically introduced damping constant. In the first term on the right hand side of Eq. (\ref{eq:FLR-model}), $P_{\mathrm{imp}}$ represents the pinning strength and $N_{\mathrm{p}}(x)=\sum_{i=1}^{N_{\mathrm{imp}}} \delta\left(x-R_i\right)$ is the distribution function of pinning sites whose positions and total number are denoted by $R_i$ and $N_{\mathrm{imp}}$, respectively, where we have assumed, for simplicity, that the pinning strength does not depend on the pinning-site position. 
In the usual case without the SAW, both $P_{\rm imp}$ and $R_i$ are static constants, and the electric field $E=E_{\rm dc}+E_{\rm ac} \, \sin\left(\omega_{\mathrm{ex}} t\right)$ gives rise to the nonlinear CDW dynamics. In the presence of the SAW, however, $P_{\rm imp}$ and $R_i$ would not be static constants any more, as explained below.

In the SAW experiment shown in the image of Fig. \ref{fig:I-V} (b), the substrate oscillation driven by the SAW with frequency $\omega_{\mathrm{ex}}$ should more or less propagate into the CDW, shaking the pinning position and strength with frequency $\omega_{\mathrm{ex}}$ \cite{FA_proceeding,C0076,SubstrateVib_pre_22}. Since the $\omega_{\mathrm{ex}}$-dependent “pinning-position” vibration turns out to play essentially the same role as $E_{\rm ac}$ \cite{Supplemental_Material, SubstrateVib_pre_22} (in \cite{C0076} treating a similar situation, due to a single-impurity assumption and a truncation of the nonlinear effect, results different from those in \cite{Supplemental_Material, SubstrateVib_pre_22} are obtained), here, we consider the $\omega_{\mathrm{ex}}$-dependent “pinning-strength” of the form $P_{\mathrm{imp}}(t)=P_{\mathrm{dc}}+P_{\mathrm{ac}} \sin \left(\omega_{\mathrm{ex}} t\right)$ which could be caused by, for example, strain-induced parameter changes \cite{Strain_Mozurkewich_prb_90} and the influences of non-uniform strains on the effective pinning \cite{Supplemental_Material}. In the presence of this temporally periodic pinning force $P_{\mathrm{imp}}(t)$, we switch on $E_{\mathrm{dc}}$ to drive the sliding motion of the CDW with $E_{\rm ac}$ remaining off. 

Since Eq. (\ref{eq:FLR-model}) cannot be solved analytically due to the nonlinear pinning force, we numerically solve Eq. (\ref{eq:FLR-model}) which can be rewritten in the dimensionless form as
\begin{equation}\label{eq:D_FLR-model}
\begin{aligned}
\frac{d\phi_{i}}{d\tilde{t}}-\left(\phi_{i+1}-2\phi_{i}+\phi_{i-1}\right)=\tilde{P}_{\mathrm{imp}} \sin (\phi_i+\beta_i)+\tilde{E}.
\end{aligned}
\end{equation}
The spatial coordinate has been discretized in units of a length scale $l$ which can be either the mean impurity distance \cite{C0003,Matsukawa_JJAP_1987} or the phase correlation length, and $\beta_{i}$ involving the randomly distributed $R_i$ is a random number between $0$ and $2\pi$ (for details, see Sec. I in \cite{Supplemental_Material}). Other dimensionless quantities are defined by $\tilde{t}=t/[(\gamma\, l^2_{\mathrm{imp}})/v^2_{\mathrm{ph}}]$, $\tilde{E}=({eQ}l^2_{\mathrm{imp}}/{m^{\ast}} v^2_{\mathrm{ph}})E$, and $\tilde{P}_{\mathrm{imp}}=(l^2_{\mathrm{imp}}/v^{2}_{\mathrm{ph}})P_{\mathrm{imp}}$. Then, the CDW current [see Eq. (\ref{CDW-current})] averaged over space and time is given by $I_{\mathrm{CDW}}\propto \frac{1}{N_{\mathrm{imp}}}\sum_{i=1}^{N_{\mathrm{imp}}} \big\langle \frac{d\phi_i}{d\tilde{t}} \big\rangle_{\tilde{t}}=\tilde{\omega}_{\phi} $,
where $\langle \rangle_{\tilde{t}}$ denotes the time average. We note that a nonlinear equation analogous to Eq. (\ref{eq:D_FLR-model}) appears in the Frenkel-Kontorova (FK) model \cite{SubstrateVib_pre_22,C0065,harmonic_paradep_Tekic_pre_07,subharmonic_paradep_Tekic_pre_11} to which the following results could be applied when the FK-potential depth vibrates.

By using the 4th-order Runge-Kutta method with a random initial configuration for $\phi_i$ and time step $\Delta \tilde{t} = 0.1$, we integrate Eq. (\ref{eq:D_FLR-model}) typically up to $\tilde{t}=2.0\times10^5$, where the first $10^4$ time steps are discarded for relaxation. Although the total number of impurities is fixed to be $N_{\rm imp}=200$, we have made spot checks that results for $N_{\rm imp}=200$ and $400$ are unchanged. The random average over the impurity distributions corresponding to the $\beta_i$ configurations is taken over 30 samples. We calculate $\tilde{\omega}_{\phi} \propto I_{\rm CDW}$ in two cases, the ac-electric-field case of $\tilde{E}=\tilde{E}_{\mathrm{dc}}+\tilde{E}_{\mathrm{ac}} \sin \left(\tilde{\omega}_{\mathrm{ex}} \tilde{t}\right)$ and $\tilde{P}_{\mathrm{imp}}=\tilde{P}_{\mathrm{dc}}$ and the SAW case of $\tilde{E}=\tilde{E}_{\mathrm{dc}}$ and $\tilde{P}_{\mathrm{imp}}=\tilde{P}_{\mathrm{dc}}+\tilde{P}_{\mathrm{ac}} \sin \left(\tilde{\omega}_{\mathrm{ex}} \tilde{t}\right)$. The former is mainly for reference. Throughout this paper, $\tilde{P}_{\rm dc}=2.0$ and $\tilde{\omega}_{\rm ex}=0.6$ are basically used.

\begin{figure}[t]
\begin{center}
\includegraphics[width=\columnwidth]{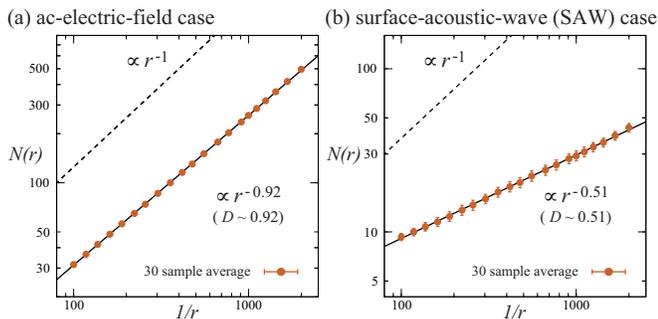}
\caption{The log-log plot of $N(r)$ as a function of $1/r$ in the (a) ac-electric-field and (b) SAW cases, where the same parameters as those in Fig. \ref{fig:I-V} are used. Solid lines indicate power-law functions of the from $(1/r)^D$ with (a) $D \sim 0.92$ and (b) $D \sim 0.51$ which are obtained by fitting the numerical data. For comparison, the slope in the trivial case of $D=1$ is also shown.
\label{fig:FLR_FD}}
\end{center}
\end{figure}

Figure \ref{fig:I-V} shows a typical example of the $\tilde{E}_{\mathrm{dc}}$ dependence of $\tilde{\omega}_{\mathrm{\phi}}$, i.e., the $I$-$V$ characteristic, for a fixed $\beta_i$ configuration in (a) the ac-electric-field case of $\tilde{E}_{\mathrm{ac}}=3.0$ and $\tilde{P}_{\mathrm{ac}}=0$ and (b) the SAW case of $\tilde{E}_{\mathrm{ac}}=0$ and $\tilde{P}_{\mathrm{ac}}=1.2$. As readily seen from the main panel of Fig. \ref{fig:I-V} (b), the Shapiro steps appear in the SAW case, as in the well-known case of the ac electric field shown in Fig. \ref{fig:I-V} (a). The harmonic and subharmonic steps can be identified from the relation $\tilde{\omega}_{\phi}=(p/q) \, \tilde{\omega}_{\rm ex}$. 
In the case of Fig. \ref{fig:I-V} where the external frequency is fixed to be $\tilde{\omega}_{\mathrm{ex}}=0.6$, the plateau at $\tilde{\omega}_{\phi} = 0.6$ corresponds to the $1/1$ harmonic step of $p = 1$ and $q = 1$, and subharmonic steps with non-integer values of $p/q$ such as the $1/2$ step of $p = 1$ and $q = 2$ corresponding to $\tilde{\omega}_{\mathrm{ex}}=0.3$ can be identified in the same manner. We note that the $0/1$ harmonic step corresponds to the non-sliding region of $\tilde{\omega}_{\phi}=0$. In each of Figs. \ref{fig:I-V} (a) and (b), consecutive subharmonic steps in a small $\tilde{E}_{\rm dc}$ window (see the magnified view shown in the inset) construct a structure similar to the entire staircase over the wide $\tilde{E}_{\rm dc}$ region. A fractal structure characterized by self-similarity of this kind is known as the devil's staircase.

To check the fractal nature of the Shapiro steps, we calculate the fractal dimension $D$ in the same manner as that in \cite{C0036_C0017,C0017_2}. In the $\tilde{E}_{\rm dc}$ region between the $0/1$ and $1/1$ steps (suppose that $l$ is the width of this region), we first count the total width $S(r)$ of steps that are larger than an arbitrarily taken step-width $r$, and then, calculate a function $N(r)=[l-S(r)]/r$ which, for the devil's staircase, should behave as $N(r)\propto(1/r)^{D}$ in the $r \rightarrow 0$ limit. We note that in the absence of the subharmonic steps as in the single impurity model, the fractal dimension $D$ is trivially 1 and that the deviation of the $D$ value from 1 points to the emergence of the fractal nature in the staircase involving the subharmonic steps.
Figure \ref{fig:FLR_FD} shows the log-log plots of $N(r)$ as a function of $1/r$ in the (a) ac-electric-field and (b) SAW cases, where in counting $S(r)$, we have regarded a plateau whose $\tilde{\omega}_{\phi}$ is unchanged within a precision of the order of $10^{-4}$ as a single Shapiro step.
As one can see from Fig. \ref{fig:FLR_FD}, the $N(r)$'s in both cases linearly increase toward $\frac{1}{r}\rightarrow \infty$ ($r \rightarrow 0$) in the log-log plot, suggestive of the fractal behavior $N(r)\propto(1/r)^{D}$. Actually, the numerical data can be well fitted by power-law functions of the form $a \, (1/r)^{D}$, where the fractal dimensions are obtained as $D\sim0.92$ and $D\sim0.51$ in the ac-electric-field and SAW cases, respectively. Although the $D$ values depend on the system parameters as reported for a similar model \cite{C0065}, the former is close to the experimental value of $D\sim 0.91$ obtained for the ac electric field \cite{C0014} and the theoretical universal value of $D=0.87$ for the circle map \cite{C0036_C0017,C0017_2}. The latter value of $D \sim 0.51$, on the other hand, is much smaller than the two above values, indicating that the mechanisms of the step formation, i.e., the mode locking, in the ac-electric-field and SAW cases are different. To see the possible difference in the mode locking likely relevant to the fractal nature, we next examine the parameter dependence of the Shapiro-step width.

\begin{figure}[t]
\begin{center}
\includegraphics[width=\columnwidth]{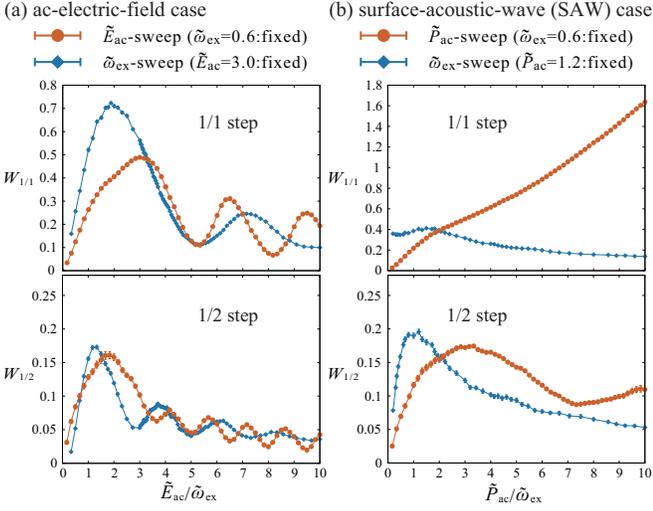}
\caption{
The parameter dependence of the harmonic $1/1$-step width $W_{1/1}$ (upper panels) and the subharmonic $1/2$-step width $W_{1/2}$ (lower panels) in the (a) ac-electric-field and (b) the SAW cases. In (a) [(b)], the horizontal axis denotes $\tilde{E}_{\rm ac}/\tilde{\omega}_{\rm ex}$ ($\tilde{P}_{\rm ac}/\tilde{\omega}_{\rm ex}$), and red and blue symbols represent the $\tilde{E}_{\mathrm{ac}}$ ($\tilde{P}_{\mathrm{ac}}$) dependence at $\tilde{\omega}_{\mathrm{ex}}=0.6$ and the $\tilde{\omega}_{\mathrm{ex}}$ dependence at $\tilde{E}_{\mathrm{ac}}=3.0$ ($\tilde{P}_{\mathrm{ac}}=1.2$), respectively. }
\label{fig:Step-width}
\end{center}
\end{figure}

Figures \ref{fig:Step-width} (a) and (b) show the parameter dependence of the $1/1$ harmonic-step width $W_{1/1}$ and the $1/2$ subharmonic one $W_{1/2}$ in the ac-electric-field and SAW cases, respectively ($W_{0/1}$ data are also available in \cite{Supplemental_Material}), where red (blue) symbols are obtained by changing the amplitude (frequency) of the time-varying external field with the frequency (amplitude) being fixed. Although the larger $\tilde{P}_{\rm ac}$ region of $\tilde{P}_{\rm ac}>\tilde{P}_{\rm dc}$ might be unrealistic, we have presented the data just for comparison with Fig. \ref{fig:Step-width} (a). In the ac-electric-field case shown in Fig. \ref{fig:Step-width} (a), the harmonic step width $W_{1/1}$ (see the upper panel) exhibits a damping oscillation with increasing $\tilde{E}_{\rm ac}$ or $1/\tilde{\omega}_{\rm ex}$ as reported elsewhere \cite{C0025,Et-Shapiro_Thorne_prb_87,C0027,C0058, harmonic_paradep_Tekic_pre_07}, and such a situation is also the case for the subharmonic-step width \cite{subharmonic_paradep_Tekic_pre_11} (see the lower panel). 
In the SAW case shown in Fig. \ref{fig:Step-width} (b), on the other hand, $W_{1/1}$ monotonically increases with increasing the amplitude $\tilde{P}_{\mathrm{ac}}$, whereas it is almost independent of the external frequency $\tilde{\omega}_{\mathrm{ex}}$, suggesting that the $1/1$ step is robust against higher-frequency SAW. The subharmonic step width $W_{1/2}$ also does not show an oscillating behavior as a function of $\frac{\tilde{P}_{\rm ac}}{\tilde{\omega}_{\mathrm{ex}}}$.
The difference can intuitively be understood by a washboard description where the swing of a particle between potential local maxima is essential for the oscillation valleys of $W_{1/1}$ \cite{C0058}. Since the swing is caused by the periodic driving force $E_{\rm ac}$, $W_{1/1}$ shows no oscillation in the SAW case of $E_{\rm ac}=0$ where the SAW corresponds to a vibration of the potential depth, i.e., a vertical motion in the washboard description, and such a situation should also be the case for other non-driving vertical vibrations.

To further examine how the CDW mode $\omega_{\phi}$ is coupled to the external frequency $\omega_{\rm ex}$, we perform the perturbative calculation proposed in \cite{C0040_C0041_C0042} (for details of the following calculation, see Sec. IV in \cite{Supplemental_Material}).
Assuming that the CDW phase takes the form of $\phi(x,t)=\phi_0(t)+\delta \phi(x,t)$ with the globally sliding mode $\phi_0(t)=\omega_\phi \, t - \frac{\overline{E}_{\rm ac}}{ \omega_{\rm ex}}\cos(\omega_{\rm ex} \, t)$ and local deviation $\delta \phi(x,t)$, we self-consistently determine $\omega_{\phi}$. For later convenience, the electric field $E$ is normalized as $\overline{E}= \big( \frac{e Q}{\gamma m^\ast} \big) \, E$. By substituting the above expression for $\phi(x,t)$ into Eq. (\ref{eq:FLR-model}), we obtain
\begin{equation}\label{eq:delta_phi}
 \delta \phi (x,t) = \frac{1}{v_{\rm ph}^2}\int dx'dt' \tilde{G}(x-x',t-t') E_{\rm P}(x', \phi(x',t')) ,
\end{equation}
\begin{equation}\label{eq:self-consistent}
\omega_{\phi} = \frac{1}{\gamma}\big\langle E_{\rm P}(x, \phi(x,t)) \big\rangle_{x,t} + \overline{E}_{\rm dc}, \\
\end{equation}
with the pinning term
\begin{equation} \label{eq:E_P}
E_{\rm P}(x, \phi(x,t)) = P_{\rm imp} N_{\rm p}(x) \sin\big( \phi_0(t) +\delta\phi(x,t)+Qx \big),
\end{equation}
where $\langle \rangle_{x,t}$ denotes the average over space and time and $\tilde{G}(x,t)=G(x,t)-\langle G(x,t) \rangle_{x,t}$ with the Green's function $G$ satisfying $\big(\frac{\gamma}{v_{\rm ph}^2} \frac{\partial}{\partial t}-\nabla^2\big)G(x,t)=\delta(x)\delta(t)$. As $E_{\rm P}(x,\phi(x,t))$ involves $\omega_{\phi}$ via $\phi_0(t)$, Eq. (\ref{eq:self-consistent}) turns out to be the self-consistent equation for $\omega_\phi$.
By using the expansion with respect to $\delta \phi$, $E_{\rm P} = P_{\rm imp} N_{\rm p}(x)\big\{  \sin\big( \phi_0(t) +Qx \big)  +   \cos\big( \phi_0(t) +Qx \big) \delta \phi + \cdots \big\}$, one can solve Eq. (\ref{eq:delta_phi}) successively to obtain $\delta \phi (x,t)$ which will further be substituted into Eq. (\ref{eq:self-consistent}) to determine $\omega_\phi$. Noting that $N_{\rm P}(x)$ represents the random impurity distribution, the leading-order contribution to $\big\langle E_{\rm P}(x, \phi(x,t)) \big\rangle_{x,t}$ turns out to be of second order in $P_{\rm imp}$, and is given by
\begin{eqnarray}\label{eq:E_P_2nd}
E_{\rm P}^{(2)}(\omega_{\phi}) &\propto& \int dt \, dt' P_{\rm imp}(t) P_{\rm imp} (t') \tilde{G}(0, t-t') \nonumber\\
&& \qquad \qquad \times \sin\big( \phi_0(t) - \phi_0(t')\big).
\end{eqnarray}

In the ac-electric-field case of $E_{\rm ac}\neq 0$ and $P_{\rm imp}=P_{\rm dc}$, $\omega_{\phi}$ and $\omega_{\rm ex}$ are coupled in the $\sin\big( \phi_0(t) - \phi_0(t')\big)$ part in Eq. (\ref{eq:E_P_2nd}) which contains
\begin{equation}\label{eq:ACE_couple}
e^{i\phi_0(t)}=e^{i[\omega_\phi t -\frac{\overline{E}_{\rm ac}}{\omega_{\rm ex}} \cos(\omega_{\rm ex}  t)]} =  \sum_{p}  (-i)^p J_p\left(\frac{\overline{E}_{\rm ac}}{ \omega_{\rm ex}} \right)  e^{i(\omega_\phi- p \omega_{\rm ex}) t  }
\end{equation}
with the Bessel function $J_p(x)$. Thus, we have
\begin{equation}\label{eq:second_order_EAC}
E_{\rm P}^{(2)}(\omega_{\phi}) \propto P_{\mathrm{dc}}^2 \sum_{p} J_p^2\left(\frac{\overline{E}_{\mathrm{ac}}}{\omega_{\mathrm{ex}}}\right){\rm Im}\big[ \tilde{H}\left(\omega_{\phi}-\textit{p}\, \omega_{\mathrm{ex}}\right)\big] ,
\end{equation}
where $\tilde{H}(\omega) = \int \frac{dk}{2\pi} \tilde{G}(k,\omega)$ is a function diverging at $\omega=0$ \cite{Matsukawa_JJAP_1987,Supplemental_Material}. Since $E_{\rm P}^{(2)}(\omega_{\phi})$ increases to diverge at $\omega_{\phi}-\textit{p}\, \omega_{\mathrm{ex}} =0$, the solution of Eq. (\ref{eq:self-consistent}) is definitely $\omega_\phi = p \, \omega_{\rm ex}$, which corresponds to the $p/1$ harmonic step. Since the coefficient $P_{\mathrm{dc}}^2 \sum_{p} J_p^2\left(\frac{\overline{E}_{\mathrm{ac}}}{\omega_{\mathrm{ex}}}\right)$ is related to the robustness of the solution against $\overline{E}_{\rm dc}$, it should determine the step width, as it can be inferred from the Bessel-function-like oscillating behavior of the step width as a function of $\frac{\overline{E}_{\mathrm{ac}}}{\omega_{\mathrm{ex}}}$ [see Fig. \ref{fig:Step-width} (a)]. We note that higher-order contributions are relevant to subharmonic steps \cite{Matsukawa_JJAP_1987}.

In the SAW case of $E_{\rm ac}=0$ and $P_{\rm imp}=P_{\rm dc}+P_{\rm ac}\sin(\omega_{\rm ex}t)$, $P_{\rm imp}(t)P_{\rm imp}(t')\sin\big( \phi_0(t) - \phi_0(t')\big)$ in Eq. (\ref{eq:E_P_2nd}) yields the $\omega_\phi$-$\omega_{\rm ex}$ coupling of the form
\begin{equation} \label{eq:SAW_couple}
P_{\rm imp}(t)e^{i\phi_0(t)}= P_{\rm dc} e^{i\omega_\phi t}+ \frac{P_{\rm ac}}{2i}\big[ e^{i(\omega_\phi + \omega_{\rm ex})t}-e^{i(\omega_\phi - \omega_{\rm ex})t} \big].
 \end{equation}
In contrast to the ac-electric-field case where the global vibration of the CDW [the oscillating part in $\phi_0(t)$] is indirectly coupled to $\omega_{\phi}$ via the pinning site [see Eq. (\ref{eq:ACE_couple})], the pinning-site vibration of the SAW directly acts on $\omega_\phi$ [see Eq. (\ref{eq:SAW_couple})]. 
Then, $E_{\rm P}^{(2)}(\omega_{\phi})$ is calculated as
\begin{equation}\label{eq:second_order_SAW}
\begin{aligned}
E_{\rm P}^{(2)}(\omega_{\phi}) \propto&{\rm Im}\left[P_{\mathrm{dc}}^2 \tilde{H}\left(\omega_{\mathrm{\phi}}\right)+\frac{P_{\mathrm{ac}}^2}{4}  \sum_{p=\pm 1}  \tilde{H}\left(\omega_{\mathrm{\phi}}
-p \, \omega_{\mathrm{ex}}\right) \right] .
\end{aligned}
\end{equation}
Due to the direct $\omega_\phi$-$\omega_{\rm ex}$ coupling, $\omega_{\rm ex}$ appears only in $\tilde{H}(\omega)$ yielding the mode-locking condition for the $1/1$ step, and its step-width determined by the coefficient of $\tilde{H}(\omega)$ becomes an $\omega_{\rm ex}$-independent increasing function of $P_{\mathrm{ac}}$, being consistent with the numerical result shown in the top panel of Fig. \ref{fig:Step-width} (b). The direct coupling process of Eq. (\ref{eq:SAW_couple}) works also in higher-order contributions relevant to other steps including the subharmonic ones, so that the Bessel-function-like oscillating behavior does not appear in the SAW case. The analytical result presented here, i.e., Eq. (\ref{eq:second_order_SAW}) [Eq. (\ref{eq:second_order_EAC})], is consistent with the numerical result shown in Fig. \ref{fig:Step-width}(b) [Fig. \ref{fig:Step-width}(a)], which suggests that the mode-locking in the SAW case (the ac-electric-field case) is direct (indirect).
Such a qualitative difference in the step formation, i.e., whether the mode locking is direct or indirect, could also affect the entire structure of the staircase, eventually leading to the difference in the fractal dimension as demonstrated in Fig. \ref{fig:FLR_FD}.

In this work, we have investigated the effect of the SAW on the overdamped sliding motion of the CDW, assuming that the SAW affects the CDW via pinning sites, where importantly, the pinning-strength vibration $P_{\mathrm{ac}}$ induces the unconventional direct mode-locking distinct from the $E_{\mathrm{ac}}$-induced indirect one. In the associated experiments, the impurity-position vibration \cite{Supplemental_Material, C0076,SubstrateVib_pre_22} may also be relevant, and the SAW generated in the piezoelectric substrate may inversely yield an electric field by the piezoelectric effect. In addition, the SAW frequency of the order of a few GHz is below but not so far from values typical of the underdamped CDW motion \cite{C0030_C0044,Underdamp_Zettl_prb_82, Underdamp_Sridhar_prl_85, Underdamp_Reagor_prb_86,Underdamp_Sridhar_prb_86} in which a $\frac{\partial^2}{\partial t^2} \phi$ term dropped in Eq. (\ref{eq:FLR-model}) becomes important.
These elements which are not taken into account in this work might be relevant to the CDW dynamics in the presence of the SAW, but experimental data enabling us to discuss them are yet to be reported.
Although the validity of the simplified modeling used here should carefully be assessed by analyzing the fractal dimension and the parameter dependence of the step width in future experimental works, we believe that this work presenting the unconventional mode-locking mechanism will promote the exploration of new classes of fractal phenomena and periodically driven systems.

\begin{acknowledgments}
The authors thank Y. Niimi and K. Fujiwara for stimulating discussions, and H. Matsukawa, H. Fukuyama, and M. Mori for valuable comments and discussions. This work is partially supported by JSPS KAKENHI Grant No. JP21K03469 and JP23H00257.
\end{acknowledgments}

\pagebreak
\widetext

\begin{center}
\large\textbf{Supplemental Material for “Fractal and subharmonic responses driven by surface acoustic waves during charge density wave sliding”}
\end{center}

\makeatletter
\renewcommand{\bibnumfmt}[1]{[S#1]}
\renewcommand{\citenumfont}[1]{S#1}


\maketitle

\section{Origin of the pinning-strength vibration}
\begin{figure}[t]
\begin{center}
\includegraphics[width=0.8\columnwidth]{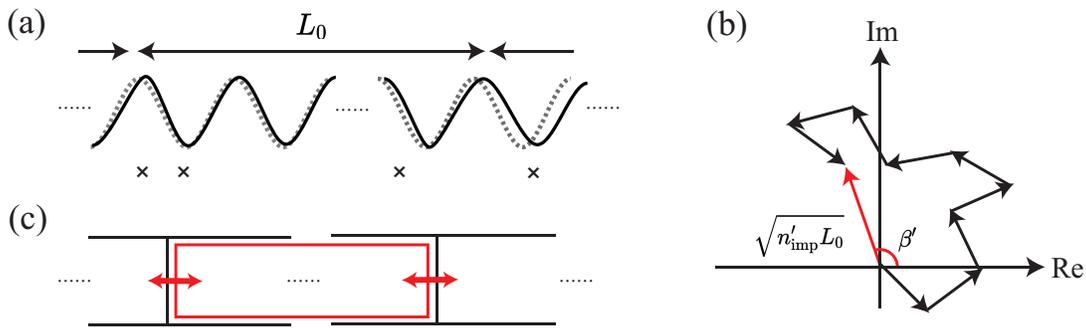}
\caption{(a) The schematically-drawn CDW state in the weak pinning regime where the original sinusoidal CDW modulation (gray dotted curve) is distorted (black solid curve) by the pinning effect of impurities (crosses). Within the FLR length $L_0$, the CDW phase is kept coherently. (b) A correspondence between ${\sum_i}^{\prime} \mathrm{e}^{i Q R_i}$ and random walk. (c) An image of time-dependent local volume changes caused by the SAW, where the characteristic volume size represented by a red box should be related to the wavelength of the SAW.}
\label{fig:random_walk}
\end{center}
\end{figure}

In this section, we discuss influences of the surface acoustic wave (SAW) on the pinning effect, providing a microscopic physical picture of the pinning strength vibration. We start from a pinning Hamiltonian for an incommensurate CDW which is given by 
\begin{equation}\label{eqS:H_pin}
H_{\mathrm{pin}}=\int dx \rho_1 V_0 \sum_{i} \delta\left(x-R_i\right)\cos (\phi(x,t)+Qx),
\end{equation}
where $R_i$ denotes the position of the $i$-th randomly-distributed impurity (pinning) site, $V_0$ is the short-range impurity potential, and other notations are the same as those of Eq. (\ref{charge-density}) in the main text ($\rho_1$ and $Q$ are the amplitude and wave number of the CDW modulation, respectively). In the strong-pinning regime, the phase at the pinning site $\phi({\bf R}_i,t)$ takes the optimal value of $\beta_i=QR_i$ for $\rho_1 V_0  <0$. The optimal value $\beta_i$ takes a random number between 0 and $2\pi$, reflecting the random distribution of $R_i$. In the weak-pinning regime where $\phi({\bf R}_i,t)$ does not necessarily take the optimal value, the CDW phase coherence is kept over the distance $L_0$ larger than the mean impurity distance $l_{\rm imp}$ [see Fig. \ref{fig:random_walk} (a)]. $L_0$ corresponding to the phase-phase correlation length is called Fukuyama-Lee-Rice (FLR) length.
In the latter case of the weak pinning, one can estimate the ''effective'' pinning defined in the spatial interval $(x-\frac{L_0}{2}, x+\frac{L_0}{2})$ in which $\phi(x,t)$ is kept coherently and thus, can be assumed to take a certain constant value $\phi'$. Following \cite{S_Fukuyama}, we perform a kind of coarse-graining to evaluate the effective pinning. By averaging Eq. (\ref{eqS:H_pin}) in the interval $(x-\frac{L_0}{2}, x+\frac{L_0}{2})$, we obtain
\begin{equation}\label{eqS:random_pin}
\begin{aligned}
\rho_1 V_0\frac{1}{L_0} {\sum_i}^{\prime}  \cos \left(\phi\left(R_i,t\right)+Q R_i\right)&=\rho_1 V_0 \frac{1}{L_0}\operatorname{Re~}\Big[ \mathrm{e}^{i \phi^{\prime}} {\sum_i}^{\prime} \mathrm{e}^{i Q R_i} \Big] \\& =  \rho_1V_0 \sqrt{n_{\mathrm{imp}}^{\prime} /L_0} \cos (\phi^{\prime}+\beta^{\prime}),
\end{aligned}
\end{equation}
where ${\sum_i}^{\prime}$ denotes the summation over $R_i$ sitting in the interval $(x - \frac{L_0}{2}, x + \frac{L_0}{2})$, and $\phi'$ and $n_{\mathrm{imp}}^{\prime}$ are a mean value of the phase and the number density of impurities in this interval. In deriving the final expression, we have used the relation
${\sum_i}^{\prime}\mathrm{e}^{i Q R_i}=\sqrt{n_{\mathrm{imp}}^{\prime} L_0} \, \mathrm{e}^{i \beta^{\prime}}$
which can be obtained by regarding $ {\sum_i}^{\prime} \mathrm{e}^{i Q R_i}$ as  random walks of $n'_{\rm imp} L_0$ steps in the complex-number plane [see Ref. \cite{S_Fukuyama} and Fig. \ref{fig:random_walk} (b)]. Thus, in the weak pinning regime, the effective pinning strength and the optimal value of the phase are given by $\rho_1 V_0\sqrt{n_{\mathrm{imp}}^{\prime} /L_0}$ and $\beta'$, respectively. It should be noted that both pinning parameters depend on the impurity distribution. The pinning strength $\rho_1 V_0 \sqrt{n_{\mathrm{imp}}^{\prime} /L_0}$ is directly related to the local impurity density $n'_{\rm imp}$, i.e, the number of impurities within the domain of length $L_0$, and the optimal value of the phase $\beta'$ takes a $[0, 2\pi]$ random number similarly to $\beta_i=QR_i$. A pure crystal sample of a CDW conductor is considered to be in the weak-pinning regime, as it is actually the case for NbSe$_3$ where $L_0$ is estimated to be approximately 1 $\mu$m \cite{S_FLR-length_Sweetland_prl_90, S_C0027}.

Now that the fundamental aspects of the pinning effect are understood, we shall consider the effects of the mechanical vibration with its frequency $\omega_{\rm ex}$ on the pinning parameters. To our knowledge, there are two associated experiments on the CDW Shapiro steps; one uses a time-dependent strain with $\omega_{\rm ex}$ of the order of MHz \cite{S_C0056} and the other uses the SAW with $\omega_{\rm ex}$ of the order of GHz \cite{S_private}. Supposing that the velocity of the acoustic wave is unchanged, a characteristic length scale in the former is 10$^3$ times larger than the wavelength of the SAW which is approximately 1 $\mu$m in the case of Ref. \cite{S_private}.  Thus, for a CDW sample of length of, for example, about 10 $\mu$m, the former should serve as a spatially uniform strain, whereas the latter SAW as a non-uniform strain. Of course, in real materials, there would exist more or less inhomogeneous sectors due to imperfection of the sample. In this sense, the following scenario with its foundation on the spatially non-uniform strains characteristic of the SAW might also be applicable to the former. Bearing these situations in our mind, we will discuss how the temporally-periodic vibrations affects the pinning parameters.

Concerning the main issue of this section, i.e., the origin of the pinning strength vibration, we have two possible scenarios. One is microscopic mechanisms such as the retardation of the conduction-electron screening of the vibrating impurity potential $V_0$ and strain-induced changes in $\rho_1$ and $Q$ characterizing the CDW state \cite{S_Strain_Mozurkewich_prb_90}. The other is an influence of spatially inhomogeneous strains on the “effective” pinning. The latter is characteristic of the SAW where the strain should be spatially non-uniform as explained above.

We first discuss the microscopic mechanisms. Naively thinking, when the CDW sample is mechanically shaken or subject to strains, impurity sites should vibrate as well. In contrast to usual metallic states where the change in the impurity potential caused by the displacement of the impurities should immediately be screened out by free electrons, in the CDW state, such a screening effect should be relatively weak as most electrons participate in the CDW condensate, so that the retardation of the screening could occur, resulting in the time dependence in $V_0$. There is another possibility that vibration-induced strains may affect the structure of the CDW, modifying  $\rho_1$ and $Q$ \cite{S_Strain_Mozurkewich_prb_90} (according to Ref. \cite{S_Strain_Mozurkewich_prb_90}, the change in $Q$ seems to be more relevant at least to an elastic softening). In the two cases, the system parameter $V_0$ or $\rho_1$ is modified periodically in time, and thus, the net pinning strength $P_{\rm imp} = \rho_1V_0$ becomes time-dependent with its frequency being basically the same as that of the mechanical vibration $\omega_{\rm ex}$.

The other scenario for the origin of the $\omega_{\rm ex}$-dependent pinning-strength is a modulation in the “effective” pinning derived in Eq. (\ref{eqS:random_pin}) in this supplemental material. In the case of the SAW, as explained above, the strain acting on the CDW sample should be spatially inhomogeneous and resultantly, local expansions and contractions should occur. Noting that the “effective” pinning strength is proportional to $\sqrt{n_{\mathrm{imp}}^{\prime}/L_0}$ and the optimal value of the phase $\beta'$ also depends on the impurity concentration, such a temporally-varying local volume change should yield the time dependence in the pinning strength and optimal value via a time dependence in $n_{\rm imp}'$.

Whichever scenario is adopted, the pinning strength ($P_{\rm imp}= \rho_1 V_0$ or $P_{\rm imp}=\rho_1 V_0 \sqrt{n_{\mathrm{imp}}^{\prime}/L_0}$) as well as the optimal local value of the phase ($QR_i$ or $\beta'$) should exhibit an oscillation with its frequency $\omega_{\rm ex}$. Thus, the pinning strength $P_{\rm imp}$ in Eq. (\ref{eq:D_FLR-model}) in the main text could be written as $P_{\rm imp}=P_{\mathrm{dc}}+P_{\mathrm{ac}}\sin(\omega_{\mathrm{ex}}t)$ as assumed in this work. We note that  the equation of motion for the CDW dynamics takes essentially the same discretized form [see Eq. (\ref{eq:D_FLR-model}) in the main text], being irrespective of whether we adopt the former or latter picture. In the former interpretation of $P_{\rm imp}= \rho_1 V_0$, the CDW phase $\phi$ should be uniform at least near each impurity site and thus, the spatial coordinate is discretized in units of the mean impurity distance $l_{\rm imp}$, whereas in the latter interpretation of $P_{\rm imp}= \rho_1 V_0 \sqrt{n_{\mathrm{imp}}^{\prime}/L_0}$, the CDW phase $\phi$ is almost uniform over the length scale $L_0$ and thus, the discretization is done in units of $L_0$. The interpretation of the length scale does not matter for the form of the pinning strength vibration $P_{\rm imp}=P_{\mathrm{dc}}+P_{\mathrm{ac}}\sin(\omega_{\mathrm{ex}}t)$. Here, the spatial uniformity in $P_{\rm imp}$ or the synchronization of the pinning-strength vibration over the whole sample is assumed for simplicity, similarly to the spatially uniform assumption for $V_0$.

\section{Effects of impurity-position vibration on the CDW sliding}
\begin{figure}[t]
\begin{center}
\includegraphics[width=0.8\columnwidth]{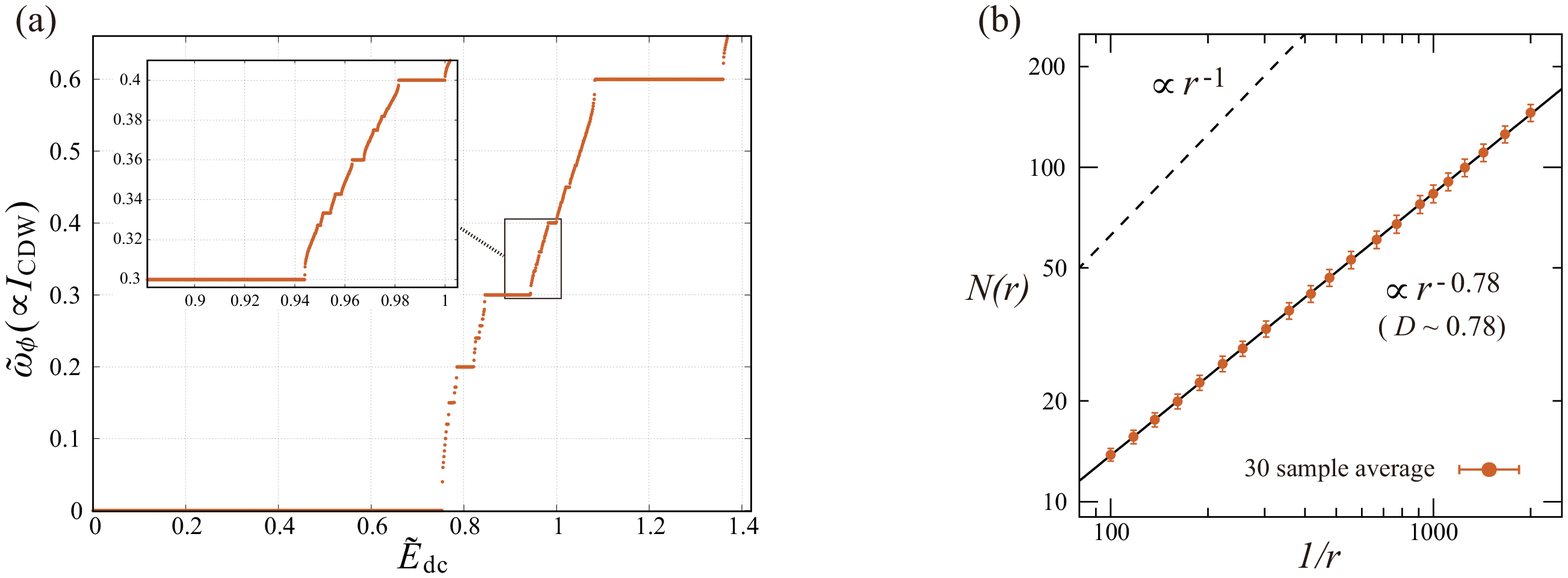}
\caption{Numerical results obtained in the case of the impurity-position vibration ($\beta_i$ vibration), where $\beta_{\mathrm{ac}}= 0.5$, $\tilde{k}=10$, $\tilde{E}_{\mathrm{ac}} = 0$, $\tilde{P}_{\mathrm{dc}} = 2.0$, $\tilde{P}_{\mathrm{ac}} = 0$, $\tilde{\omega}_{\rm ex}=0.6$, and $N_{\mathrm{imp}} = 200$ are used. (a) The $\,\tilde{E}_{\mathrm{dc}}$ dependence of $\tilde{\omega}_{\phi}\,$ (the $I$-$V$ characteristics) and (b) the log-log plot of $N(r)$ as a function of $1/r$, where the figure notations are the same as those in Figs. \ref{fig:I-V} and \ref{fig:FLR_FD} in the main text.}
\label{fig:B_IV_FD}
\end{center}
\end{figure}
In the main text, the time-dependent pinning strength of the form $P_{\mathrm{imp}}(t)=P_{\mathrm{dc}}+P_{\mathrm{ac}} \sin \left(\omega_{\mathrm{ex}} t\right)$ is considered as a typical simplified model for the SAW applied on the substrate. On the other hand, a mechanical vibration of the SAW may vibrate the impurity-position (or the optimal value of the phase at pinning site) with the SAW frequency $\omega_{\mathrm{ex}}$. Such a SAW-induced position shift from the mean impurity position $R_i$ could be written as $\delta x_i=A \sin \left(\omega_{\mathrm{ex}} t+k R_i\right)$, where $k$ is the wave number of the SAW and $A$ should be proportional to the amplitude of the SAW. Then, the pinning term in Eq. (\ref{eq:FLR-model}) in the main text reads
\begin{equation}\label{eqS:supple_1}
\begin{aligned}
P_{\mathrm{imp}} N_{\mathrm{p}}(x)\sin (\phi+Qx)=P_{\mathrm{imp}} \sum_{i=1}^{N_{\mathrm{imp}}} \sin \Big(\phi+QR_i + QA \sin(\omega_{\mathrm{ex}} t+k R_i ) \Big),
\end{aligned}
\end{equation}
where $\delta x_i$ has been incorporated in Eq. (\ref{eqS:supple_1}) via $N_{\mathrm{p}}(x)=\sum_{i=1}^{N_{\mathrm{imp}}} \delta\left(x-R_i-\delta x_i\right)$.
The associated dimensionless equation of motion in discretized form becomes
\begin{equation}\label{eqS:FLR-model_beta}
\begin{aligned}
\frac{d\phi_{i}}{d\tilde{t}}-\left(\phi_{i+1}-2\phi_{i}+\phi_{i-1}\right)=\tilde{P}_{\mathrm{imp}}\sin \Big(\phi_i+\beta_{i,\mathrm{dc}} + \beta_{\mathrm{ac}}\sin \left( \tilde{\omega}_{\mathrm{ex}} \tilde{t} + \tilde{k}\beta_{i,\mathrm{dc}} \right) \Big)+\tilde{E},
\end{aligned}
\end{equation}
where $\beta_{i,\mathrm{dc}}=Q R_i$ is a random number between $0$ and $2\pi$, and the dimensionless amplitude and wave number of the SAW are denoted as $\beta_{\mathrm{ac}}=AQ$ and $\tilde{k}=k/Q$, respectively. By comparing the above equation (\ref{eqS:FLR-model_beta}) and Eq, (\ref{eq:D_FLR-model}) in the main text, one notices that the impurity-position vibration serves as a temporal oscillation in the random number $\beta_i$ in the pinning term.

Figure \ref{fig:B_IV_FD} (a) shows the $\tilde{E}_{\mathrm{dc}}$ dependence of the CDW frequency $\tilde{\omega}_\phi$, i.e., the $I$-$V$ characteristics, for the $\beta_i$ vibration of $\beta_{\mathrm{ac}}= 0.5$ and $\tilde{k}=10$, where the pinning strength and the electric field are assumed to be static, namely, $\tilde{P}_{\rm ac}=0$ and $\tilde{E}_{\rm ac}=0$, and other parameters are the same as those in the main text. As readily seen in Fig. \ref{fig:B_IV_FD} (a), Shapiro steps with a self-similar structure appear as in the cases of the ac electric field and the time-dependent pinning strength (see Fig. \ref{fig:I-V} in the main text). The occurrence of the harmonic and subharmonic steps in the presence of the substrate vibration of this kind has been reported in the similar Frenkel-Kontorova (FK) model \cite{S_SubstrateVib_pre_22} where $\beta_i$ in our model is fixed to be $\beta_i=0$. It is shown in Ref. \cite{S_SubstrateVib_pre_22} that the Shapiro-step width exhibits a damping oscillation as a function of the vibration amplitude similarly to the conventional ac-electric-field-induced one. According to Ref. \cite{S_C0076} where the associated one-degree-of-freedom model corresponding to $N_{\rm imp}=1$ in Eq. (\ref{eqS:FLR-model_beta}) is considered, the difference between the SAW and the ac electric field seems to be reflected in higher harmonic steps (subharmonic steps do not appear due to the lack of multiple degrees of freedom), although in Ref. \cite{S_C0076}, the nonlinear pinning effect is linearized with respect to the vibration amplitude and as a result, the phases of the 0th and 1st order pinning terms are shifted exactly by $\pi$, possibly causing an accidental cancellation. Indeed, the 2/1 harmonic step which is reported to disappear for $\lambda=0$ in \cite{S_C0076} survives in the corresponding multi-impurity FK model without such a truncation \cite{S_SubstrateVib_pre_22}.

To quantitatively evaluate the self-similar structure shown in Fig. \ref{fig:B_IV_FD} (a), we calculate the fractal dimension $D$ in the same manner as that in obtaining Fig. \ref{fig:FLR_FD} in the main text. Figure \ref{fig:B_IV_FD} (b) shows $N(r)$ as a function of $1/r$ for thse same parameters as those for Fig. \ref{fig:B_IV_FD} (a), where $N(r)$ is defined in the main text. By fitting the numerical data with a power-law function with exponent $D$, the fractal dimension $D$ is obtained as $D \sim 0.78$ for the parameter set used here. The obtained value $D \sim 0.78$ is larger than the fractal dimension for the pinning-strength vibration $D\sim 0.51$, but is rather close to the universal value of $D=0.87$ expected for the ac electric field (see Fig. \ref{fig:FLR_FD} and the associated discussion in the main text).

The fractal dimension and the parameter dependence of the step width suggest that the impurity-position vibration plays a role similar to that of the ac electric field, which can be understood in the following way. In Eq. (\ref{eqS:FLR-model_beta}), the external frequency $\omega_{\rm ex}$ appears in the form of $\sin\big( X+ \sin(\omega_{\rm ex} t) \big)$. As discussed in the main text, the $\omega_{\rm ex}$ dependence of this kind is essential for the indirect mode-locking working in the ac-electric-field case, so that the essentially same mode-locking mechanism is naively expected for the impurity-position vibration.

%


\section{Parameter dependence of the 0/1-step width}
\begin{figure}[t]
\begin{center}
\includegraphics[width=0.8\columnwidth]{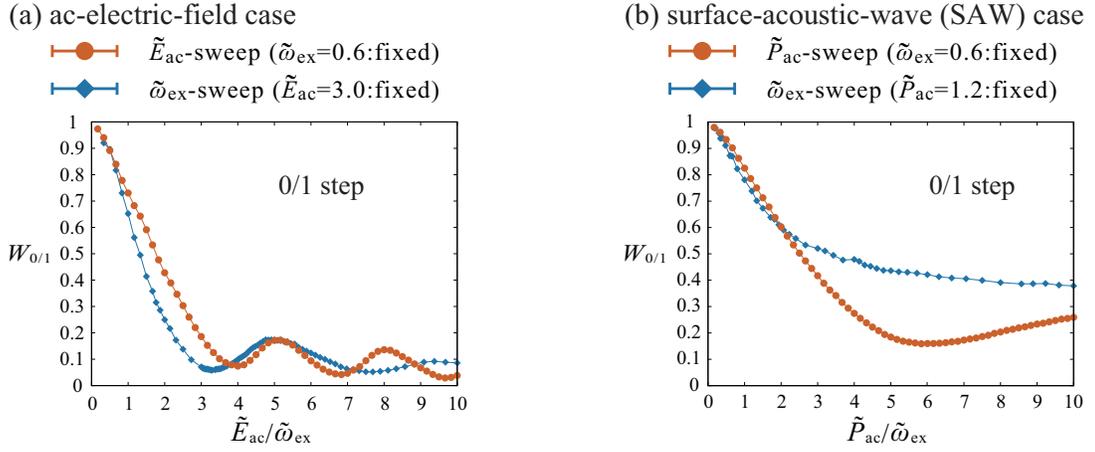}
\caption{
The parameter dependence of the $0/1$-step width $W_{0/1}$ (threshold field)  in the (a) ac-electric-field and (b) SAW cases, where the parameter values and color notations are the same as those in Fig. 3 in the main text.}
\label{fig:Step-width-0/1}
\end{center}
\end{figure}

In the main text, we discuss the parameter dependence of the widths of the 1/1 harmonic-step and the 1/2 subharmonic-step as typical examples of the staircase [see Fig. \ref{fig:Step-width} in the main text]. In this supplemental material, for completeness, we show the associated results for the 0/1 harmonic-step width $W_{0/1}$ corresponding to the threshold field in Fig. \ref{fig:Step-width-0/1}.
In the ac-electric-field case shown in Fig. \ref{fig:Step-width-0/1} (a), $W_{0/1}$ exhibits a damping oscillation with increasing $\tilde{E}_{\rm ac}$ or $1/\tilde{\omega}_{\rm ex}$ as reported in Ref. \cite{S_Et-Shapiro_Thorne_prb_87}. In the SAW case shown in Fig. \ref{fig:Step-width-0/1} (b), on the other hand, $W_{0/1}$ gets suppressed with increasing $\tilde{P}_{\rm ac}$ or $1/\tilde{\omega}_{\rm ex}$ without showing an oscillating behavior. We note that as commented in the main text, although the larger $\tilde{P}_{\rm ac}$ region of $\tilde{P}_{\rm ac}>\tilde{P}_{\rm dc}$ ($\tilde{P}_{\rm ac}/\tilde{\omega}_{\rm ex} > 3.3 $ for the parameter set used here) would be unrealistic, the data in this larger $\tilde{P}_{\rm ac}$ region are presented just for comparison with the Fig. \ref{fig:Step-width-0/1} (a).  

\section{Perturbative analysis of the mode-locking phenomenon}
In the main text, the mode-locking mechanisms in the ac-electric-field and SAW cases are discussed based on the perturbative analysis. Although the perturbative method in the ac-electric-field case has already been reported in Ref. \cite{S_C0040_C0041_C0042} and can straightforwardly be extended to the SAW case, here, we provide the details of the calculation for completeness.

For later convenience, we introduce $\overline{\gamma} \equiv \gamma / v_{\mathrm{ph}}^2$, $\overline{E}_{\mathrm{P}} \equiv \left(P_{\mathrm{imp}}(t) /v_{\mathrm{ph}}^2 \right)N_{\mathrm{p}}(x)\sin (\phi+Qx)$, and $\overline{E} \equiv \left({eQ}/{m^{\ast} \gamma}\right) E$, and rewrite Eq. (\ref{eq:FLR-model}) in the main text as follows:
\begin{equation}\label{eqS:FLR-model}
\begin{aligned}
\left(\overline{\gamma}\frac{\partial}{\partial t}-\nabla^2\right) \phi (x,t)=\overline{E}_{\mathrm{P}}(x,t)+ \overline{\gamma} \overline{E}(t).
\end{aligned}
\end{equation}
In the sliding regime, we could express the CDW phase $\phi(x,t)$ as $\phi(x, t)= \phi_0(t)+ \delta \phi(x,t)$ with a globally sliding mode $\phi_0(t)$ given by
\begin{equation}\label{eqS:global}
\phi_0(t)=\omega_{\phi} t - \dfrac{\overline{E}_{\mathrm{ac}}}{\omega_{\mathrm{ex}}}\cos\left(\omega_{\mathrm{ex}}t\right)
\end{equation}
and the local deviation from it $\delta \phi(x,t)$, where we have assumed that $\phi_0(t)$ takes the same form as that of the solution of Eq. (\ref{eqS:FLR-model}) without the pinning term. It should be emphasized that $\omega_{\phi}$ of our interest, which is proportional to the CDW sliding current $I_{\rm CDW}$, is not a given parameter but is to be determined taking the pinning effect into account. To derive the equation for $\omega_\phi$, we first take the spatial average of Eq. (\ref{eqS:FLR-model}) to obtain
\begin{equation}\label{eqS:FLR-space}
\overline{\gamma}\langle\dot{\phi}(x, t)\rangle_x=\left\langle \overline{E}_{\mathrm{P}}(x,t)\right\rangle_x + \overline{\gamma}\Big(\overline{E}_{\mathrm{dc}}+\overline{E}_{\mathrm{ac}} \sin(\omega_{\rm ex} t) \Big) ,
\end{equation}
where the surface term has been dropped.
By further taking the time average of Eq. (\ref{eqS:FLR-space}), we have
\begin{equation}\label{eqS:self-consistent}
\omega_{\phi}=\frac{1}{\overline{\gamma}}\left\langle \overline{E}_{\mathrm{P}}(x,t)\right\rangle_{x,t} +\overline{E}_{\mathrm{dc}},
\end{equation}
where $\langle A \rangle_x=\frac{1}{L}\int dx \, A$ and $\langle A \rangle_t=\frac{1}{T}\int dt \, A$ represent the spatial and time averages of $A$, respectively, and $\langle \dot{\delta \phi}(x,t) \rangle_{x,t}=0$ has been assumed. Since $\overline{E}_{\mathrm{P}}(x,t)$ depends on the CDW phase $\phi=\phi_0 +\delta \phi$ involving $\omega_\phi$ in $\phi_0$, Eq. (\ref{eqS:self-consistent}) turns out to be a self-consistent equation for $\omega_{\phi}$. On the other hand, $\overline{E}_{\mathrm{P}}(x,t)$ also involves $\delta \phi(x,t)$ which can be determined by
\begin{equation}\label{eqS:delta_phi}
\begin{aligned}
\delta \phi(x, t)=& \int d x^{\prime} d t^{\prime} G\left(x-x^{\prime}, t-t^{\prime}\right) \left\{\overline{E}_{\mathrm{P}}\left(x^{\prime}, t^{\prime}\right)-\langle \overline{E}_{\mathrm{P}}\left(x^{\prime},t^{\prime}\right)\rangle_{x,t}\right\} \\
=& \int d x^{\prime} d t^{\prime} \tilde{G}\left(x-x^{\prime}, t-t^{\prime}\right)  \overline{E}_{\mathrm{P}}\left(x^{\prime}, t^{\prime}\right) ,
\end{aligned}
\end{equation}
where $\tilde{G}(x, t)$ is defined by $\tilde{G}(x, t)\equiv G(x, t)-\langle G(x, t)\rangle_{x, t}$ with the Green function $G(x,t)$ satisfying
\begin{equation}
\left(\overline{\gamma} \frac{\partial}{\partial t}-\nabla^2\right) G(x, t)=\delta(x) \delta(t).
\end{equation}
Note that the Fourier transformation yields
\begin{equation}
G(k, \omega)=\left(\mathrm{i} \overline{\gamma} \omega+k^2\right)^{-1}, \qquad \tilde{G}(k, \omega) = G(k, \omega)-G(0,0) \delta(k) \delta(\omega).
\end{equation}
By using the Taylor expansion with respect to $\delta \phi$
\begin{equation}\label{eqS:EP_ex}
\begin{aligned}
\overline{E}_{\mathrm{P}}(x, t) 
& = \left( P_{\rm imp}(t) N_{\rm p}(x) /v_{\rm ph}^2 \right) \sum_{n=0}^{\infty} \frac{1}{n !} \sin\Big( \phi_0(t) +Qx+ \frac{n}{2}\pi \Big) [\delta \phi(x,t)]^n \\
& \equiv \sum_{n=0}^{\infty} \overline{E}_{\mathrm{P}}^{(n+1)}(x,t),
\end{aligned}
\end{equation}
one can solve Eq. (\ref{eqS:delta_phi}) successively as follows:
\begin{equation}
\delta \phi^{(n)}(x,t)=\int d x^{\prime} d t^{\prime} \tilde{G}\left(x-x^{\prime},t-t^{\prime}\right) \overline{E}_{\mathrm{P}}^{(n)}\left(x^{\prime},t^{\prime}\right).
\end{equation}
Here, $\overline{E}_{\mathrm{P}}^{(n+1)}(x,t)$ and $\delta \phi^{(n)}(x,t)$ denote the $n$-th order contributions in $P_{\rm imp}$. By further substituting the solution into Eq. (\ref{eqS:self-consistent}), we obtain
\begin{equation}
\omega_{\phi} = \frac{1}{\overline{\gamma}} \sum_{n=0}^{\infty} \left\langle\overline{E}_{\mathrm{P}}^{(n+1)}(x,t)\right\rangle_{x,t} +\overline{E}_{\mathrm{dc}} .
\end{equation}
Now, the problem is reduced to calculate the concrete expression of $\langle \overline{E}_{\mathrm{P}}^{(n+1)}(x,t) \rangle_{x,t}$ as a function of $\omega_\phi$ so that one can solve the self-consistent equation for $\omega_\phi$.
With the use of the following abbreviation
\begin{equation}
C_{\mathrm{P}}(x, t) \equiv \cos \left(Q x+\phi_0(t)\right), \qquad \qquad S_{\mathrm{P}}(x, t) \equiv \sin \left(Q x+\phi_0(t)\right),
\end{equation}
the first- and second-order terms in Eq. (\ref{eqS:EP_ex}) can be expressed as
\begin{equation}
\begin{aligned}
&\overline{E}_{\mathrm{P}}^{(1)}(x,t)=v_{\rm ph}^{-2}P_{\mathrm{imp}}(t) N_{\mathrm{P}}(x) S_{\mathrm{P}}(x,t), \\
&\overline{E}_{\mathrm{P}}^{(2)}(x,t)=v_{\rm ph}^{-4}\int d x_1 d t_1 P_{\rm imp}(t)P_{\rm imp}(t_1)N_{\mathrm{P}}(x)N_{\mathrm{P}}(x_1)\tilde{G}\left(x-x_1,t-t_1\right) C_{\mathrm{P}}(x,t) S_{\mathrm{P}}\left(x_1,t_1\right) .
\end{aligned}
\end{equation}
After the spatial average, we have
\begin{eqnarray}
\left\langle \overline{E}_{\mathrm{P}}^{(1)}(x,t)\right\rangle_x &=& v_{\rm ph}^{-2}P_{\mathrm{imp}}(t)\left\langle N_{\mathrm{P}}(x)S_{\mathrm{P}}(x, t)\right\rangle_x = \frac{1}{v_{\rm ph}^{2}L} P_{\mathrm{imp}}(t) \sum_{i=1}^{N_{\mathrm{imp}}} \sin \left(Q R_i+\phi_0(t)\right), \nonumber\\
\left\langle \overline{E}_{\mathrm{P}}^{(2)}(x,t)\right\rangle_x &=& \frac{1}{v_{\rm ph}^{4}L}\sum_{i,j=1}^{N_{\mathrm{imp}}}  \int d t_1 P_{\rm imp}(t)P_{\rm imp}(t_1)\tilde{G}\left(R_i-R_j, t-t_1\right) C_{\mathrm{P}}\left(R_i, t\right) S_{\mathrm{P}}\left(R_j, t_1\right) \nonumber\\
&=&\frac{1}{v_{\rm ph}^{4}L}\sum_{i,j=1}^{N_{\mathrm{imp}}}  \int d t_1 P_{\rm imp}(t)P_{\rm imp}(t_1)\int\frac{dk}{2\pi}\tilde{G}\left(k, t-t_1\right) \, e^{ik(R_i-R_j)} C_{\mathrm{P}}\left(R_i, t\right) S_{\mathrm{P}}\left(R_j, t_1\right).
\end{eqnarray}
Since the pinning sites $R_{i}$ are randomly distributed, $\frac{1}{L}\sum_{i=1}^{N_{\mathrm{imp}}} f(R_i)$ serves as a random average of a function $f(x)$. Thus, $\langle \overline{E}_{\mathrm{P}}^{(1)}(x,t)\rangle_x$ vanishes, as the random average of $\sin \left(Q R_i+\phi_0(t)\right)$ vanishes, i.e., $\frac{1}{L}\sum_{i=1}^{N_{\rm imp}} e^{iQR_i}=0$.
The second-order term $\langle \overline{E}_{\mathrm{P}}^{(2)}(x,t)\rangle_x$, on the other hand, becomes nonvanishing after the random average in the specific cases of $R_j = R_i$, which can be understood from the fact that a component in 
\begin{equation}
e^{ik(R_i-R_j)}C_{\mathrm{P}}\left(R_i, t\right) S_{\mathrm{P}}\left(R_j, t_1\right) = \frac{e^{ik(R_i-R_j)}}{2} \Big[ \sin\Big( Q(R_i-R_j)+\phi_0(t)-\phi_0(t_1) \Big) + \sin\Big( Q(R_j+R_i) + \phi_0(t_1) + \phi_0(t)\Big) \Big]
\end{equation}
becomes $R_i$-independent for $R_j = R_i$. The nonvanishing contribution coming from $R_j=R_i$ turns out to be
\begin{equation}\label{eqS:AC_s}
\begin{aligned}
\left\langle \overline{E}_{\mathrm{P}}^{(2)}(x,t)\right\rangle_x &=\frac{1}{v_{\rm ph}^{4}L} \sum_{i=1}^{N_{\mathrm{imp}}}\int d t_1 P_{\rm imp}(t)P_{\rm imp}(t_1)\int\frac{dk}{2\pi}\tilde{G}\left(k, t-t_1\right) e^{ik(R_i-R_i)} \frac{1}{2} \sin\big( Q(R_i-R_i)+\phi_0(t)-\phi_0(t_1) \big), \\
\end{aligned}
\end{equation}
and its time-average  is given by
\begin{eqnarray}\label{eqS:AC_s2}
&&\left\langle \overline{E}_{\mathrm{P}}^{(2)}(x,t)\right\rangle_{x,t} =\frac{1}{T} \int dt \, \frac{N_{\text {imp}}}{v_{\rm ph}^{4}L}\int d t_1 P_{\rm imp}(t)P_{\rm imp}(t_1)\int\frac{dk}{2\pi}\tilde{G}\left(k, t-t_1\right)\frac{1}{4i}\left[e^{i\left\{\phi_0(t)-\phi_0(t_1)\right\}}-e^{-i\left\{\phi_0(t)-\phi_0(t_1)\right\}}\right] \nonumber\\
&&\quad \propto \int \frac{d k}{2\pi} \int d \omega \int dt \, d t_1 P_{\rm imp}(t)P_{\rm imp}(t_1)\tilde{G}(k, \omega)e^{-i\omega(t-t_1)} \frac{1}{4i}\left[e^{i\left\{\omega_{\phi}\left(t-t_1\right)-\frac{\overline{E}_{\mathrm{ac}}}{\omega_{\mathrm{ex}}}\big(\cos( \omega_{\mathrm{ex}} t)-\cos  (\omega_{\mathrm{ex}} t_1) \big)\right\}}-\mathrm{c.c.}\right],
\end{eqnarray}
where Eq. (\ref{eqS:global}) has been used.

In the ac-electric-field case of $E_{\rm ac}\neq 0$ and $P_{\rm imp}(t)=P_{\rm dc}$, Eq. (\ref{eqS:AC_s2}) reads
\begin{eqnarray}\label{eqS:AC_s3}
\left\langle \overline{E}_{\mathrm{P}}^{(2)}(x,t)\right\rangle_{x,t} &\propto& P_{\rm dc}^2 \int \frac{d k}{2\pi}\int d \omega \, \tilde{G}(k, \omega) \\
&&\times  \int dt \, d t_1  \, \frac{1}{4i}\left[e^{i\left\{\left(-\omega+\omega_{\phi}\right)\left(t-t_1\right)-\frac{\overline{E}_{\mathrm{ac}}}{\omega_{\mathrm{ex}}}\big(\cos( \omega_{\mathrm{ex}} t)-\cos  (\omega_{\mathrm{ex}} t_1) \big)\right\}}-e^{i\left\{\left(-\omega-\omega_{\phi}\right)\left(t-t_1\right)+\frac{\overline{E}_{\mathrm{ac}}}{\omega_{\mathrm{ex}}}\big(\cos( \omega_{\mathrm{ex}} t)-\cos  (\omega_{\mathrm{ex}} t_1) \big)\right\}}\right]. \nonumber
\end{eqnarray}
By using the formula for the Bessel function of the 1st kind $J_p(x)$ ($p$:integer)
\begin{equation}\label{eqS:ACE_couple}
e^{\pm i\frac{\overline{E}_{\rm ac}}{\omega_{\rm ex}} \cos(\omega_{\rm ex}  t)} =  \sum_{p}  (\pm i)^p J_p\big(\frac{\overline{E}_{\rm ac}}{ \omega_{\rm ex}} \big)  e^{\pm i p \, \omega_{\rm ex} \, t  },
\end{equation}
one can perform the integration over time in Eq. (\ref{eqS:AC_s3}) as follows:
\begin{eqnarray}\label{eqS:AC_s4}
\left\langle \overline{E}_{\mathrm{P}}^{(2)}(x,t)\right\rangle_{x,t} &\propto& P_{\rm dc}^2 \int \frac{d k}{2\pi}\int d \omega \, \tilde{G}(k, \omega)\sum_{p,q}  (-i)^p i^q J_p\big(\frac{\overline{E}_{\rm ac}}{ \omega_{\rm ex}} \big)J_q\big(\frac{\overline{E}_{\rm ac}}{ \omega_{\rm ex}} \big)\nonumber\\
&\times& \int dt \, d t_1 \, \frac{1}{4i} \left[e^{i\left(-\omega+\omega_{\phi}-p\, \omega_{\rm ex}\right) t} e^{-i\left(-\omega+\omega_{\phi}-q\, \omega_{\rm ex}\right) t_1}-e^{i\left(-\omega-\omega_{\phi}+q\, \omega_{\rm ex} \right)t} e^{-i\left(-\omega-\omega_{\phi}+p\, \omega_{\rm ex} \right)t_1}\right] \nonumber\\
&=& P_{\rm dc}^2 \sum_{p}  J^2_p\big(\frac{\overline{E}_{\rm ac}}{ \omega_{\rm ex}} \big) \int \frac{d k}{2\pi} \frac{1}{4i}\Big[ \tilde{G}(k, \omega_\phi-p\, \omega_{\rm ex}) - \tilde{G}(k, -\omega_\phi+ p\, \omega_{\rm ex})  \Big] \nonumber\\
&=& \dfrac{P_{\mathrm{dc}}^2}{2} \sum_p J_p^2\left(\frac{\overline{E}_{\mathrm{ac}}}{\omega_{\mathrm{ex}}}\right) \operatorname{Im}\left\{\tilde{H}\left(\omega_\phi-\textit{p}\, \omega_{\mathrm{ex}}\right)\right\}
\end{eqnarray}
Here, $\tilde{H}(\omega)$ is defined by $\tilde{H}(\omega) \equiv \int\frac{d k}{2 \pi}\tilde{G}(k, \omega)  \equiv  \mathcal{F}^{(+)}(\omega)+i\, \mathcal{F}^{(-)}(\omega)$, where $\mathcal{F}^{(\pm)}(\omega)$ is calculated as
\begin{equation}\label{eqS:H}
\mathcal{F}^{(\pm)}(\omega)=\frac{(\overline{\gamma} \omega)^{-1 / 2}}{4 \sqrt{2} \pi}\left\{f_l(\omega) \pm 2 f_t\left(\frac{\sqrt{2 \overline{\gamma} \omega} \Lambda}{\overline{\gamma} \omega-\Lambda^2}\right)\right\}
\end{equation}
with a cutoff $\Lambda$ and
\begin{equation}\label{eqS:f}
\begin{aligned}
f_l(x) & \equiv \log \left|\frac{\Lambda^2-\sqrt{2 \overline{\gamma} x} \Lambda+\overline{\gamma} x}{\Lambda^2+\sqrt{2 \overline{\gamma} x} \Lambda+\overline{\gamma} x}\right|, \qquad
f_t(x)&\equiv
\left\{
\begin{alignedat}{2}
& \tan^{-1} x &\quad & x \geq 0 \\
& \pi+\tan ^{-1} x &\quad & x<0.
\end{alignedat}
\right.
\end{aligned}
\end{equation}
One can see from Eq. (\ref{eqS:H}) that each component of $\tilde{H}(\omega)$, i.e., $\mathcal{F}^{(\pm)}(\omega)$, diverges at $\omega=0$. Since $\langle \overline{E}_{\mathrm{P}}^{(2)}(x,t)\rangle_{x,t}$ diverges at $\omega_\phi = p \, \omega_{\rm ex}$ [see the final expression in Eq. (\ref{eqS:AC_s4})], the solution of the self-consistent equation (\ref{eqS:self-consistent}) is definitely $\omega_\phi = p \, \omega_{\rm ex}$ which corresponding to the mode-locking condition for the harmonic Shapiro steps. The subharmonic Shapiro steps occur as higher-order contributions $\langle \overline{E}_{\mathrm{P}}^{(n)}(x,t)\rangle_{x,t}$ with $n \geq 4$ \cite{S_C0040_C0041_C0042}. Further details of this perturbative approach can be found in Ref. \cite{S_C0040_C0041_C0042}.

In the SAW case of $E_{\rm ac}=0$ and $P_{\mathrm{imp}}(t)=P_{\mathrm{dc}}+P_{\mathrm{ac}} \sin \left(\omega_{\mathrm{ex}} t\right)$, the vibration part in  $\langle \overline{E}_{\mathrm{P}}^{(2)}(x,t)\rangle_{x,t}$ can be expressed as a simple linear combination of the Fourier modes, i.e.,
\begin{equation}
P_{\text {imp }}(t) P_{\mathrm{imp}}\left(t_1\right)=P_{\rm dc}^2 + \frac{P_{\rm dc}P_{\rm ac}}{2i}\big( e^{i\omega_{\rm ex}t}-e^{-i\omega_{\rm ex}t} + e^{i\omega_{\rm ex}t_1}-e^{-i\omega_{\rm ex}t_1} \big)-\frac{P_{\rm ac}^2}{4} \big( e^{i\omega_{\rm ex}(t+t_1)}-e^{-i\omega_{\rm ex}(t-t_1)} - e^{i\omega_{\rm ex}(t-t_1)}+e^{-i\omega_{\rm ex}(t+t_1)} \big), \nonumber
\end{equation}
which is in sharp contrast to the ac-electric-field case of Eq. (\ref{eqS:ACE_couple}) involving the Bessel function in the Fourier-mode expansion. Then, the second-order contribution is calculated as
\begin{eqnarray}\label{eqS:SAW_s}
&&\left\langle \overline{E}_{\mathrm{P}}^{(2)}(x,t)\right\rangle_{x,t} \propto \int \frac{d k}{2\pi} \int d \omega \, \tilde{G}(k, \omega) \int dt d t_1 \, \bigg[ \frac{P_{\rm dc}^2}{4i} \left[ e^{-i(\omega-\omega_\phi)\left(t-t_1\right)}-e^{-i(\omega+\omega_\phi)\left(t-t_1\right)}\right] \nonumber\\
&&- \frac{P_{\rm dc}P_{\rm ac}}{8}\left[ \left\{ \big(e^{-i(\omega-\omega_{\rm ex}-\omega_\phi)t} - e^{-i(\omega+\omega_{\rm ex}-\omega_\phi)t}\big) e^{i(\omega-\omega_\phi)t_1} - \big(e^{-i(\omega-\omega_{\rm ex}+\omega_\phi)t} - e^{-i(\omega+\omega_{\rm ex}+\omega_\phi)t}\big) e^{i(\omega+\omega_\phi)t_1}  \right\} - \Big\{ t \leftrightarrow -t_1 \Big\} \right] \nonumber\\
&&+\frac{P_{\rm ac}^2}{16i} \Big[ e^{-i(\omega+\omega_{\rm ex}-\omega_\phi)(t-t_1)}-e^{-i(\omega+\omega_{\rm ex}+\omega_\phi)(t-t_1)}+e^{-i(\omega-\omega_{\rm ex}-\omega_\phi)(t-t_1)}-e^{-i(\omega-\omega_{\rm ex}+\omega_\phi)(t-t_1)} \nonumber\\
&& \qquad \qquad \qquad +\Big\{ e^{-i(\omega-\omega_{\rm ex}+\omega_\phi)t} e^{i(\omega+\omega_{\rm ex}+\omega_\phi)t_1}-  e^{-i(\omega-\omega_{\rm ex}-\omega_\phi)t} e^{i(\omega+\omega_{\rm ex}-\omega_\phi)t_1} \Big\}  + \Big\{ t \leftrightarrow -t_1 \Big\} \Big] \bigg] \nonumber\\
&=& \int \frac{d k}{2\pi} \int d \omega \, \tilde{G}(k, \omega) \frac{1}{4i}\bigg[ \Big\{ P_{\rm dc}^2 \delta(\omega-\omega_\phi) +\frac{P_{\rm ac}^2}{4} \Big(\delta(\omega-\omega_\phi+\omega_{\rm ex})+ \delta(\omega-\omega_\phi-\omega_{\rm ex})\Big)\Big\} - \Big\{ \omega \rightarrow -\omega \Big\} \bigg] \nonumber\\
&=& \frac{1}{2}\left[P^2_{\mathrm{dc}}\mathrm{Im}\left\{\tilde{H}(\omega_\phi)\right\}+\frac{P^2_{\mathrm{ac}}}{4}\mathrm{Im}\left\{\tilde{H}(\omega_\phi-\omega_{\mathrm{ex}})+\tilde{H}(\omega_\phi+\omega_{\mathrm{ex}})\right\}\right].
\end{eqnarray}
As in the ac-electric-field case, the mode-locking occurs at $\omega=0$ of $\tilde{H}(\omega)$, so that in the SAW case, the $0/1$ and $1/1$ harmonic steps corresponding to $\omega_\phi=0$ and $\omega_\phi=\omega_{\rm ex}$, respectively, appear within the second-order perturbation [see the final expression in Eq. (\ref{eqS:SAW_s})]. Higher-order contributions yield the mode-locking conditions for other steps including the subharmonic ones. Even in that case, the Bessel function never appears due to the direct coupling between $\omega_{\rm ex}$ and $\omega_\phi$.

\end{document}